\documentclass[prx,reprint,twocolumn,amsmath,superscriptaddress,longbibliography]{revtex4-2}
\usepackage{amsmath,amssymb,bm,mathrsfs,graphicx, braket, times,amsthm,enumerate, scrextend}
\usepackage[colorlinks=true,citecolor=blue,linkcolor=blue]{hyperref}
\usepackage{longtable}
\usepackage{multirow}
\usepackage{array}
\usepackage{bigstrut}
\usepackage[all,cmtip]{xy}
\usepackage[normalem]{ulem}
\usepackage[usenames,dvipsnames]{color}
\usepackage{makecell}
\usepackage{xcolor}

\begin{document}
\title{Catalog of phonon emergent particles}
\author{Dongze Fan}
\affiliation{National Laboratory of Solid State Microstructures and School of Physics, Nanjing University, Nanjing 210093, China}
\affiliation{Collaborative Innovation Center of Advanced Microstructures, Nanjing University, Nanjing 210093, China}
\affiliation{International Quantum Academy, Shenzhen 518048, China.}
\author{Hoi Chun Po}
\affiliation{Department of Physics, Hong Kong University of Science and Technology, Clear Water Bay, Hong Kong, China}
\author{Xiangang Wan}
\affiliation{National Laboratory of Solid State Microstructures and School of Physics, Nanjing University, Nanjing 210093, China}
\affiliation{Collaborative Innovation Center of Advanced Microstructures, Nanjing University, Nanjing 210093, China}
\affiliation{Hefei National Laboratory, Hefei 230088, China}
\author{Feng Tang}\email{fengtang@nju.edu.cn}
\affiliation{National Laboratory of Solid State Microstructures and School of Physics, Nanjing University, Nanjing 210093, China}
\affiliation{Collaborative Innovation Center of Advanced Microstructures, Nanjing University, Nanjing 210093, China}

\begin{abstract}
The outcome of conventional topological materials prediction scheme could sensitively depend on first-principles calculations parameters. Symmetry, as a powerful tool, has been exploited to enhance the reliability of predictions. Here, we establish the relationship between the Wyckoff positions (WYPOs) and the phonon wavefunctions at each high-symmetry point (HSP) in all 230 space groups (SGs). Based on this, on one hand, we obtain a complete mapping from WYPO to the occurrence of emergent particles (EMPs) at each HSP in 230 SGs, and establish several rules of enforcing EMPs for phonons; on the other hand, we determine the contribution of the WYPO to the phonon angular momentum. Then we unambiguously identify 20,516,167 phonon EMPs in 111,872 materials in two databases. The purely symmetry-determined wavefunctions generalize the conventional Bloch theorem, could find a wide scope of application to physical properties related with basis functions of irreducible representations.

\end{abstract}
\maketitle
\date{\today}

\section{Introduction}
\label{sec1}
In the past decades, the topology of electronic band structures has attracted extensive interest \cite{Weyl-Wan, RMP-kane, RMP-Qi, AP-Black, RMP-Bansil, RMP-Chiu, RMP-AV, RMP-LV, Nature-Review-Mag}. Tens of thousands of materials with nontrivial topological electronic properties have been predicted theoretically and computationally \cite{N-4, high-science}, and some have undergone experimental verification \cite{N-1, N-2, N-3, experiment-1, experiment-2, experiment-3, experiment-4, experiment-5, experiment-6}. In addition, phonon, which describes the collective motion of crystals, are a significant type of elementary excitation within condensed matter systems and provides another platform to realize nontrivial band topology in solid materials. Phonons play an essential role in many physical properties of solid materials, including superconductivity, electrical conductivity, thermal conductivity, and heat capacity \cite{phonon-book}. Topological phonons have become a rapidly evolving new area.

The topological band crossings in condensed matter systems, such as Weyl point (WP) \cite{Weyl-Wan}, triple point (TP) \cite{triple-point-1, triple-point-2}, Dirac point (DP) \cite{Dirac-point-1,Dirac-point-2,Dirac-point-3}, where the low-energy excitation near them behaves like some emergent particles (EMPs), can exhibit various topological charges \cite{charge-1, charge-2}, pseudospin structures \cite{pseudospin-1, pseudospin-2}, and other properties. Conventionally, predicting these EMPs using the first-principles calculation is influenced by certain computational parameters. For example, in phononic system, the choices of several parameters, such as the supercell size, and $\bm{k}$ point mesh affect the numerical precision of the force constants, and could change prediction on the topological property considerably. Symmetry, underlying many elegant methods/rules, can be used to enhance the reliability of predictions, such as using the filling constraint to guarantee a topological band crossing to exist without any realistic calculations. \cite{filling-1, filling-2}.

Very recently, the study of these EMPs in phonon system has become a hot spot \cite{phonon-review, high-throu-phonon-2,high-throu-phonon-3, zhangtiantian-phonon-PRL,charge4phonon-liu,diracphonon-PRL,sixflodphonon,app-triple,chemical-2}. These EMPs could unlock additional degrees of freedom in devices engineered from topological electrons. Utilizing these EMPs has the potential to catalyze the development of a range of innovative technologies. As an example, the WP can give rise to the phonon arc, which arises from the projections of the oppositely charged WPs \cite{Weyl-Wan}. This can provide a one-way phonon propagation channel that is robust and immune to defects, and can be used to achieve the directional selectivity of conducting heat at terahertz scales. It holds potential for applications in the phonon waveguides \cite{app-waveguides-1}, abnormal heat transport \cite{app-heat-2}, and more. As another example, the TP can introduce extra scattering channels within the three phonon-phonon scattering processes, which can suppress lattice thermal conductivity, consequently improving the thermoelectric properties of the material \cite{app-triple}. Moreover, the surface states of the EMPs can enhance electron-phonon coupling \cite{en-super-1}, potentially leading to topological superconductivity at the surface \cite{en-super-2}. From a chemical perspective, the phonon surface states could promote catalysis when the phonon frequency resonates with certain intermediate steps of the reaction \cite{chemical-1,chemical-2}.

On the other hand, chiral phonon possessing nonzero angular momentum (AM) was proposed theoretically \cite{Lifa-L-1} and verified experimentally \cite{chiral-obser}. Phonon chiral introduces a new degree of freedom to the study of fundamental physics. The chiral phonons can be applied to develop spin-phononic devices and are important for energy-efficient information processing \cite{chiral-obser, phonon-computer-2, phonon-computer-3}. The recent study has uncovered that the interaction between chiral phonons and electrons significantly contributes to superconductivity, which may induce unconventional and high-temperature superconductivity, providing a new perspective for understanding high-temperature superconductivity \cite{chiral-super}. Furthermore, the research on chiral phonons has also extended to the field of biology. Chiral phonons in microcrystals and nanofibrils of biomolecules have been reported \cite{bio-chiral-2}. Recent research also revealed that the generation of the polarizing spin current within the DNA helical chains is linked to the chiral phonons \cite{bio-chiral-3}.

\begin{figure*}[t]
\centering\includegraphics[width=0.9\textwidth]{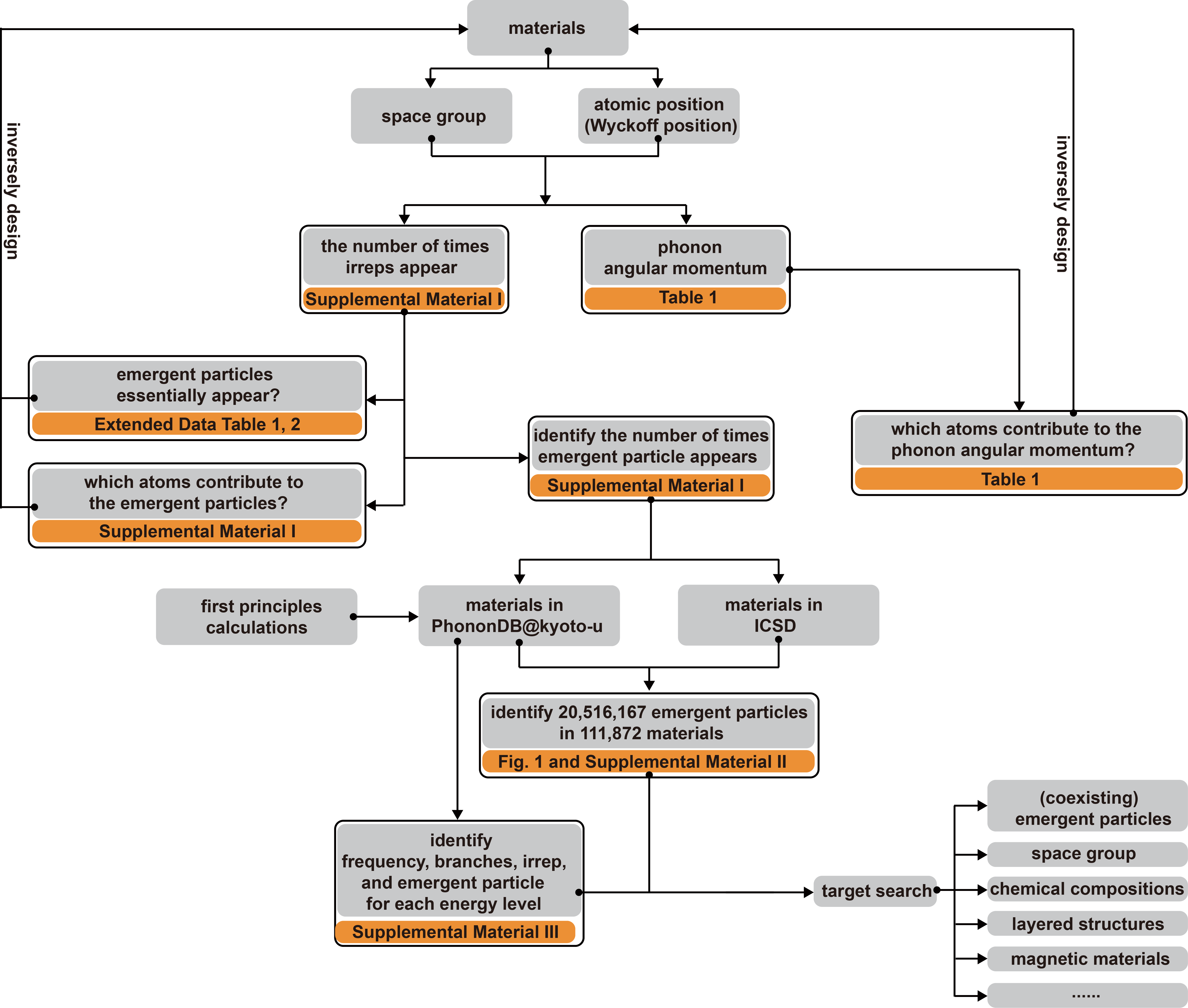}
\caption{Schematic diagram summarizing the results obtained in this work. The locations of these results are indicated in the orange boxes. Phonons possess only vibrating degrees of freedom in three directions, which can be regarded as three $p$ orbitals. Therefore, only the vector representation needs to be considered. Based on this, we obtain relationship between the WYPOs and the phonon wavefunctions. Consequently, the relationship between atomic positions and the number of occurrences of irreps, phonon AM, and other physical quantities related to the phonon wavefunctions can be obtained.}
\label{fig1}
\end{figure*}

However, until now, there are very few experimental verifications for topological phononic materials \cite{exper-phonon-1,exper-phonon-2,exper-phonon-3,Weyl-Chiral}. In response to this question, on one hand, high-throughput calculations can be performed to provide numerous subsequent materials. On the other hand, is it possible to propose a new and reliable simple approach for predicting topological phonons? Furthermore, all previously topological materials prediction schemes only exploit the number of occurences of irreducible representations (irreps). However, the origin in these irreps, rather than simply the occurrence,  has been overlooked.

Herein, we extend beyond the conventional method of symmetry analysis in phonon systems. We establish the connection between the Wyckoff positions (WYPOs) and the phonon wavefunctions at each high-symmetry point (HSP) in 230 space groups (SGs). Based on this, on one hand, it can be clearly known the number of times each irrep appears at high-symmetry points (HSPs) in the phonon spectrum, thereby obtaining the occurrences of various types of EMPs. We establish a catalog of various types of EMPs in phonon system, which contains 111,872 materials and 20,516,167 EMPs. These materials are obtained from the public phonon material database at Kyoto university (PhononDB@kyoto-u) \cite{phonopyDB} and the Inorganic Crystal Structure Database (ICSD) \cite{ICSD}. The frequencies and branches of 1,694,951 EMPs contained in 10,034 materials from the PhononDB@kyoto-u are also determined quantitatively by the force constants. We selected 2,309 nearly ideal candidate materials, which merit further experimental investigation. On the other hand, we identified the atomic positions that contribute to the phonon AM at HSPs. This also provides a new idea for the follow-up study of phonon AM. The flowchart of this work is shown in Fig. \ref{fig1}.

\begin{figure*}[t]
\centering\includegraphics[width=0.9\textwidth]{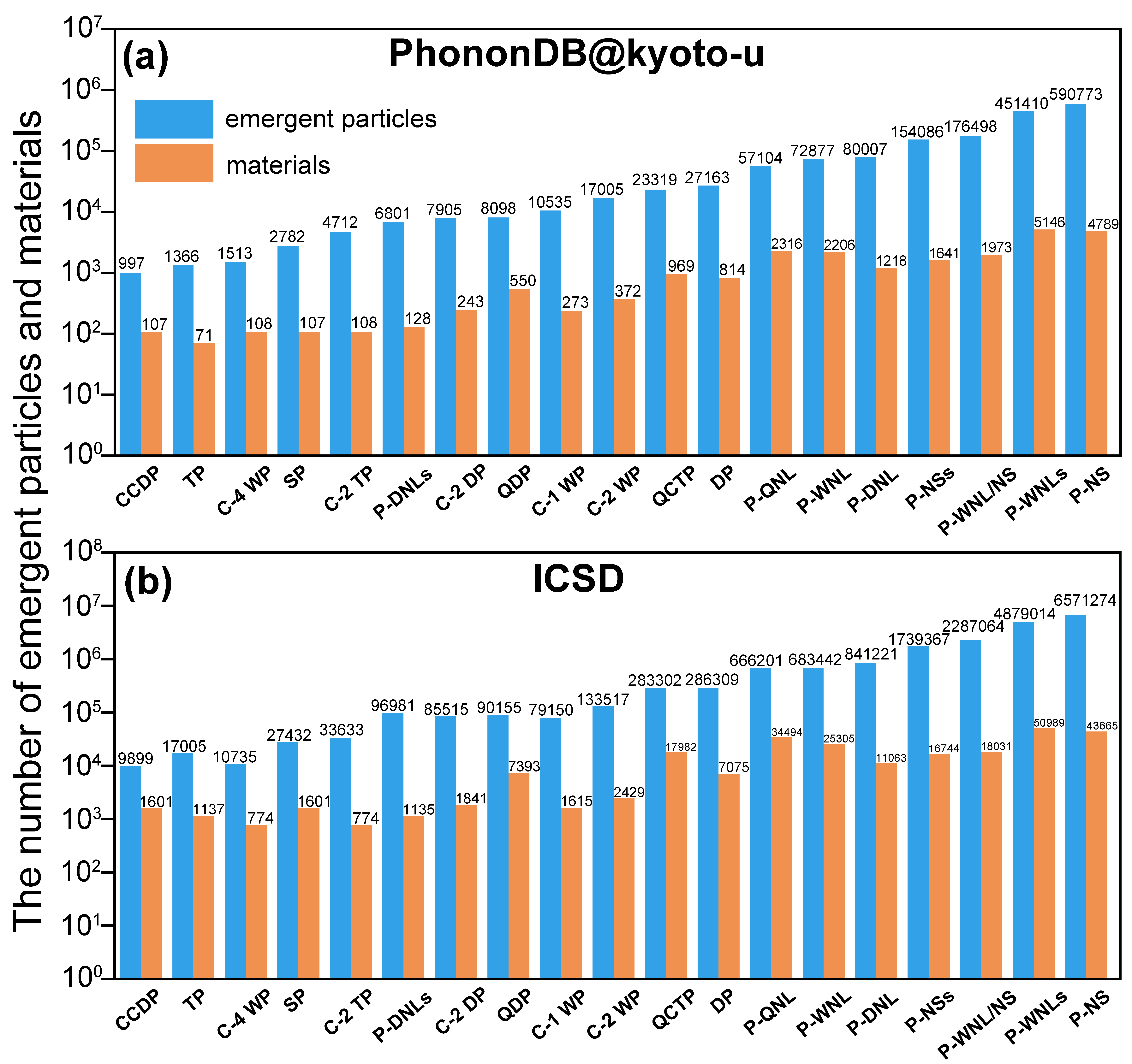}
\caption{The number of different types of EMPs, which we identified in the phonon spectra of materials from PhononDB@kyoto-u and ICSD, along with the count of the materials hosting the corresponding type of EMP. The 19 types of EMPs on the horizontal axis are all types of EMPs that can exist at the HSPs in 230 SGs in phonon system. The meanings of the abbreviations of these 19 types of EMPs are provided in Extended Data. The data of the quantity is plotted on a logarithmic scale. The blue bar represents the number of EMPs, and the orange bar represents the number of materials hosting the corresponding type of EMP. The number on each bar represents the quantity of the EMPs or materials.}
\label{fig2}
\end{figure*}

\begin{figure*}[t!]
\centering\includegraphics[width=0.9\textwidth]{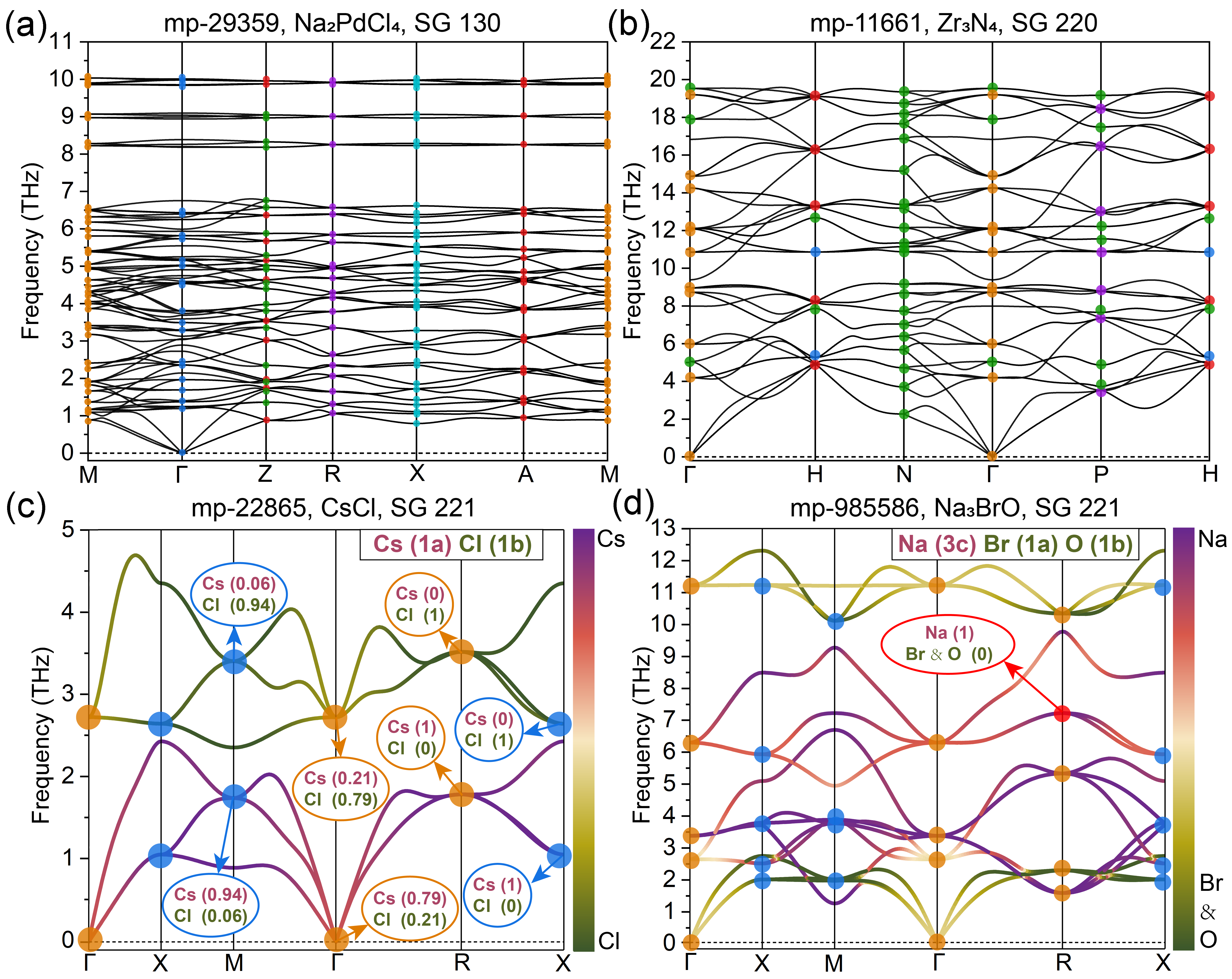}
\caption{The phonon spectra of four ideal topological phonon materials. (a) The phonon dispersion of Na$_2$PdCl$_4$ with MPID 29359 and SG 130. The DP is marked by purple circle, the QDP is marked by red circle, the P-NS is marked by cyan circle, the P-NSs is marked by orange circle, the P-QNL is marked by blue circle and the P-WMLs is marked by green circle at the HSPs. (b) The phonon dispersion of Zr$_3$N$_4$ with MPID 11661 and SG 220. The DP is marked by purple circle, the QCTP is marked by orange circle, the CCDP is marked by blue circle, the SP is marked by red circle and the P-WNLs is marked by green circle at the HSPs. (c) The projected phonon dispersion of CsCl with MPID 22865 and SG 221. The QCTP is marked by orange circle and the P-QNL is marked by blue circle at the HSPs. (d) The projected phonon dispersion of Na$_3$BrO with MPID 985586 and SG 221. The QCTP is marked by orange circle, the P-QNL is marked by blue circle and the P-WNLs is marked by red circle at the HSPs. The color bars in (c) and (d) represent the atomic contribution to the phonon vibration by the projection of the eigen displacements. In Na$_3$BrO, our focus lies on the contribution of the atoms occupying the 3c WYPO, namely Na, thus we combine the contributions of the atom at the 1a WYPO, Br, with those at the 1b WYPO, O. The deeper the purple, the greater the contribution of Cs (Na) to the phonon vibration; the deeper the green, the greater the contribution of Cl (Br and O) to the phonon vibration. In (c) and (d), the numbers inside the circles represent the contributions of different atoms to the EMPs, with the maximum value being 1 and the minimum value being 0.}
\label{fig3}
\end{figure*}

\section{Statistics of materials and emergent particles}
\label{sec2}

In this paper, we focus on the EMPs that can exist at HSP, due to the fact that EMPs occurring on high-symmetry line and high-symmetry plane are often accidental. Based on Ref. \cite{SciBul-yao}, we obtain the exhaustive list of EMP types that can be realized at HSP in phonon system. Details regarding the collection of materials, the determination of the types of EMP, and so on are included in the Extended Data.

In the PhononDB@kyoto-u, we collected 10,034 materials, of which 9,380 were identified as having EMPs at HSPs. This includes 19 elementary substances, 1,139 binary compounds, and 5,399 ternary compounds. In the ICSD, we collected 101,838 materials, of which 91,789 were identified as having EMPs at HSPs. This includes 1,865 elementary substances, 27,561 binary compounds, and 37,196 ternary compounds. It can be seen that over 90\% of these materials host EMPs at HSPs.

The statistics for the number of the EMPs and the materials hosting them in the phonon spectra, identified from PhononDB@kyoto-u and ICSD, are shown in Fig. \ref{fig2} (a) and Fig. \ref{fig2} (b). We have identified over 20 million EMPs in total. These EMPs are universally present in the phonon spectra, which enables the broad possibility of realizing topological phonons in various materials toward diverse potential applications.

All of these results are collected in Supplemental Material (SM) II and SM III. One can search for target EMPs/chemical compositions/crystal symmetry/ICSD number/Material Project IDentification number (MPID) in SM II or SM III to get concrete compounds for further studies. For example, if one wants to search for materials hosting charge-4 WP (C-4 WP), one can directly search for ``C-4 WP'' in SM II (where the abbreviations of EMPs are provided in the Extended Data) to obtain the specific information of the materials containing C-4 WP, as well as all EMPs at HSPs included in these materials, within the above two databases. If the material is in PhononDB@kyoto-u, then one can search for this material in SM III to access its phonon spectrum, as well as the specific frequencies, branches, and positions where all the EMPs appear at HSP in this material.

Recently, two dimensional (2D) materials have attracted increasing attention \cite{2Ddata,new2Ddata-1,new2Ddata-2}, particularly the chiral phonon within these materials \cite{Lifa-L-1,Lifa-L-2,Lifa-L-3,chiral-obser,chiral-gam,2Dpnonon-apply,phononAM-zong}. For example, phenomena such as the entanglement of chiral phonons and single photons \cite{chiral-gam}, as well as vibrational dichroism of chiral valley phonons \cite{2Dpnonon-apply}, have been discovered in 2D materials. Based on Ref. \cite{2Ddata}, we have identified 559 layered materials, which are included in PhononDB@kyoto-u, that can be used to obtain 2D materials through exfoliation, as shown in Extended Data Table 3. Additionally, these obtained 2D materials can further yield an additional 613 2D materials through chemical substitution. These materials can serve as candidate materials for the study of phonon AM.

\section{The definitely existing emergent particles}
\label{sec3}

The classification in Ref. \cite{SciBul-yao} takes into account the given SG, $\bm{k}$ point, and the irrep of the little group of $\bm{k}$ point, obtaining a specific EMP, thus being able to determine in which SG a target EM can appear. However, it is generally unclear whether these EMPs definitely appear in structures that meet the SG requirements. A straightforward criterion for the definitive existence of a EMP is as follows: If the occurrence of this EMP is allowed at a $\bm{k}$ point, and all the irreps of this $\bm{k}$ point correspond to this EMP, then its presence is certain. These EMPs are listed in Extended Data Table 1. We refer to these EMPs as representation-enforced EMPs. Below, employing the $n_{\bm{k},j}^{w}$, which represents the number of occurrences of the irrep $j$ at $\bm{k}$ point when the WYPO $w$ is occupied, we establish two additional  rules to enforce the definitive existence of the EMP at a $\bm{k}$ point, which strikingly reveals that nearly all EMPs at HSPs definitely appear as long as the structure belongs to the SG that allows the occurrence of this EMP for phonons.

These two rules are as follows: (1) Not all irreps correspond the same type of EMP, but this type of EMP will appear when atoms occupy any WYPO in the SG. (2) Not every WYPO contributes to the same type of EMP, however, when atoms only occupy the WYPOs that do not contribute to this type of EMP, the resulting structure will belong to another SG. We refer to the EMPs that are enforced to appear by these two rules as the EMPs enforced by WYPOs, and are listed in Extended Data Table 2.

We find that, other than P-WNLs, all EMPs at HSPs are enforced to definitely appear through exhaustively examining all WYPOs in the 230 SGs. Besides, only for SGs 216, 221, 225, 227 and 229, P-WNLs have a chance to be not present.

\begin{table*}[!t]
\caption{All the HSPs in the 20 SGs allowing nonvanishing phonon AM, along with the WYPO contributing to the phonon AM for the corresponding (co-)irrep. These WYPOs are placed within the parentheses following the (co-)irrep(s). For example, at the $K$ point of SG 187, for the phonon mode corresponding to co-irrep \{4\} to carry nonvanishing AM, the atoms should occupy at least one WYPO in \{2h, 2i, 6n, 12o\}.}\label{tab3}
   \begin{tabular}{m{0.6cm}<{\centering}m{1cm}<{\centering}p{15.4cm}}
    \hline \hline
\noalign{\vskip 4pt}
\multirow{1}{*}{SG} & \multicolumn{1}{c}{HSP} & \multicolumn{1}{c}{WYPO which contributes the nonzero phonon AM for the corresponding (co-)irreps} \\ \noalign{\vskip 4pt} \hline  
\noalign{\vskip 3pt}
44 & $W$ & \{1\}: (2b, 4c, 4d, 8e);  \enspace  \{2\}: (2a, 4c, 4d, 8e)  \\ 
\noalign{\vskip 3pt}
45 & $W$ & \{1, 1\}: (4b, 8c);  \enspace  \{2, 2\}: (4a, 8c)  \\ 
\noalign{\vskip 3pt}
46 & $W$ & \{1, 2\}: (4a, 4b, 8c)  \\ 
\noalign{\vskip 3pt}
82 & $P$ & 1, 3: (2c, 2d, 4f, 8g);  \enspace 2, 4: (2a, 2b, 4e, 8g)  \\ 
\noalign{\vskip 3pt}
119 & $P$ & \{1\}, \{3\}: (2c, 2d, 4f, 8g, 8h, 8i, 16j);  \enspace \{2\}, \{4\}: (2a, 2b, 4e, 8g, 8h, 8i, 16j)  \\
\noalign{\vskip 3pt}
120 & $P$ & \{1, 3\}: (4c, 4d, 8e, 8g, 8h, 16i);  \enspace \{2, 4\}: (4a, 4b, 8e, 8f, 8h, 16i)  \\ 
\noalign{\vskip 3pt}
143 & $K$, $H$ & 1: (1b, 1c, 3d);  \enspace 2: (1a, 1c, 3d);  \enspace 3: (1a, 1b, 3d)  \\ 
\noalign{\vskip 3pt}
144 & $K$, $H$  & 1, 2, 3: (3a)  \\ 
\noalign{\vskip 3pt}
145 & $K$, $H$ & 1, 2 ,3: (3a)  \\ 
\noalign{\vskip 3pt}
149 & $K$, $H$ & \{1\}: (1c, 1d, 1e, 1f, 2h, 2i, 3j, 3k, 6l);  \enspace  \{2\}: (1a, 1b, 1e, 1f, 2g, 2i, 3j, 3k, 6l);  \enspace  \{3\}: (1a, 1b, 1c, 1d, 2g, 2h, 3j, 3k, 6l)  \\ 
\noalign{\vskip 3pt}
151 & $K$, $H$ & \{1\}, \{2\}, \{3\}: (3a, 3b, 6c)  \\ 
\noalign{\vskip 3pt}
153 & $K$, $H$ & \{1\}, \{2\}, \{3\}: (3a, 3b, 6c)  \\ 
\noalign{\vskip 3pt}
156 & $K$, $H$ & \{1\}: (1b, 1c, 3d, 6e);  \enspace  \{2\}: (1a, 1c, 3d, 6e);  \enspace \{3\}: (1a, 1b, 3d, 6e)  \\ 
\noalign{\vskip 3pt}
158 & $K$ & \{1\}: (2b, 2c, 6d);  \enspace \{2\}: (2a, 2c, 6d);  \enspace \{3\}: (2a, 2b, 6d)  \\ 
\noalign{\vskip 3pt}
 & $H$ & \{1, 1\}: (2b, 2c, 6d);  \enspace  \{2, 2\}: (2a, 2c, 6d);  \enspace  \{3, 3\}: (2a, 2b, 6d)  \\ 
\noalign{\vskip 3pt}
174 & $K$ & 1: (1c, 1d, 1e, 1f, 2h, 2i, 3j, 3k, 6l);  \enspace 2: (2g, 2h, 6l);  \enspace  3: (1a, 1b, 1e, 1f, 2g, 2i, 3j, 3k, 6l);  \enspace  4: (2h, 2i, 6l);  \enspace 5: (1a, 1b, 1c, 1d, 2g, 2h, 3j, 3k, 6l);  \enspace 6: (2g, 2i, 6l)  \\ 
\noalign{\vskip 3pt}
 & $H$ & 1: (1c, 1e, 2h, 2i, 3j, 6l);  \enspace  2: (1b, 1d, 2g, 2h, 3k, 6l);  \enspace  3: (1a, 1e, 2g, 2i, 3j, 6l);  \enspace  4: (1d, 1f, 2h, 2i, 3k, 6l);  \enspace 5: (1a, 1c, 2g, 2h, 3j, 6l);  \enspace 6: (1b, 1f, 2g, 2i, 3k, 6l)  \\ 
\noalign{\vskip 3pt}
187 & $K$ & \{1\}: (1c, 1d, 1e, 1f, 2h, 2i, 3j, 3k, 6l, 6m, 6n, 12o);  \enspace  \{2\}: (2g, 2h, 6n, 12o);  \enspace  \{3\}: (1a, 1b, 1e, 1f, 2g, 2i, 3j, 3k, 6l, 6m, 6n, 12o);  \enspace  \{4\}: (2h, 2i, 6n, 12o);  \enspace  \{5\}: (1a, 1b, 1c, 1d, 2g, 2h, 3j, 3k, 6l, 6m, 6n, 12o); \enspace \{6\}: (2g, 2i, 6n, 12o) \\ 
\noalign{\vskip 3pt}
 & $H$ & \{1\}: (1c, 1e, 2h, 2i, 3j, 6l, 6n, 12o);  \enspace  \{2\}: (1b, 1d, 2g, 2h, 3k, 6m, 6n, 12o);  \enspace  \{3\}: (1a, 1e, 2g, 2i, 3j, 6l, 6n, 12o);  \enspace  \{4\}: (1d, 1f, 2h, 2i, 3k, 6m, 6n, 12o);  \enspace  \{5\}: (1a, 1c, 2g, 2h, 3j, 6l, 6n, 12o);  \enspace  \{6\}: (1b, 1f, 2g, 2i, 3k, 6m, 6n, 12o)  \\ 
\noalign{\vskip 3pt}
188 & $K$ & \{1\}: (2c, 2d, 2e, 2f, 4h, 4i, 6j, 6k, 12l);  \enspace  \{2\}: (2a, 2c, 4g, 4h, 6j, 12l);  \enspace  \{3\}: (2a, 2b, 2e, 2f, 4g, 4i, 6j, 6k, 12l);  \enspace  \{4\}: (2c, 2e, 4h, 4i, 6j, 12l); \enspace  \{5\}: (2a, 2b, 2c, 2d, 4g, 4h, 6j, 6k, 12l);  \enspace  \{6\}: (2a, 2e, 4g, 4i, 6j, 12l)  \\ 
\noalign{\vskip 3pt}
 & $H$ & \{1, 4\}: (2c, 2d, 2e, 2f, 4h, 4i, 6j, 6k, 12l);  \enspace  \{2, 5\}: (2a, 2b, 2c, 2d, 4g, 4h, 6j, 6k, 12l);  \enspace  \{3, 6\}: (2a, 2b, 2e, 2f, 4g, 4i, 6j, 6k, 12l)  \\ 
\noalign{\vskip 3pt}
196 & $W$ & \{1\}: (4c, 4d, 16e, 24f, 24g, 48h);  \enspace  \{2\}: (4a, 4b, 16e, 24f, 24g, 48h)  \\ 
\noalign{\vskip 3pt}
216 & $W$ & \{1\},  \{3\}: (4c, 4d, 16e, 24f, 24g, 48h, 96i);  \enspace  \{2\}, \{4\}: (4a, 4b, 16e, 24f, 24g, 48h, 96i)  \\ 
\noalign{\vskip 3pt}
219 & $W$ & \{1, 3\}: (8b, 24c, 24d, 32e, 48f, 48g, 96h);  \enspace  \{2, 4\}: (8a, 24c, 24d, 32e, 48f, 48g, 96h)  \\ \noalign{\vskip 4pt} \hline \hline
  \end{tabular}
\end{table*}

\section{Ideal topological phonon materials}
\label{sec4}

We select four ideal topological phonon materials, further illustrating the EMPs enforced by representation and WYPO, and explaining how to determine which atoms contribute to the EMPs based on our results. Their phonon spectra are shown in Fig. \ref{fig3}.

Fig. \ref{fig3} (a) shows the phonon spectrum of material Na$_2$PdCl$_4$ with SG 130. According to first-principles calculations, all branch nodes are fourfold degeneracies at the $R$ point, which correspond to DPs; and all branch nodes are twofold degeneracies at the $A$, $X$, and $M$ points, which correspond to QDPs, P-NS[s] (the plural form of P-NS), and P-NSss (the plural form of P-NSs), respectively. This is consistent with the results obtained from symmetry analysis which show that these EMPs are all enforced by representation at the above HSPs (Extended Data Table 1). In SG 130, the P-QNL at the $\Gamma$ point, and P-WNLs along with QDP at the $Z$ point are all enforced to appear by WYPO (Extended Data Table 2). Therefore, P-QNL, P-QNLs or QDP will definitely appear at the corresponding HSP, and will present along with other types of EMP or topologically trivial bands. These are in accordance with the phonon spectrum in Fig. \ref{fig3} (a). There are a large number of different types of EMPs in this material, and the bands between 8 THz and 11 THz are well separated from other bands, which facilitates the observation of these EMPs experimentally.

Then, we present the material Zr$_3$N$_4$ with SG 220, whose phonon spectrum is shown in Fig. \ref{fig3} (b). Here, we concentrate on the sixfold degenerate points, SP, which have garnered significant interest due to their maximum degeneracy in bosonic systems. The CCDP, SP, and P-WNLs are all enforced to appear at $H$ point in SG 220 by WYPO (Extended Data Table 2). At the $H$ point of this material, there exist five SPs marked by red circles, two CCDPs marked by blue circles, and two P-WNLss (the plural form of P-WNLs) marked by green circles, and there are no topologically trivial bands. In the high-frequency part of phonon spectrum, these SPs are well separated, which facilitates observing them in experiments.

Next, we use an example to illustrate the limitations of predicting topological phonons based on all representations of the little groups of the $\bm{k}$ point, which is the conventional method for predicting topological phonons. The co-irreps corresponding to P-WNLs at HSPs in SG 221 are: \{5\} and \{6\} at $\Gamma$ point, and \{5\} and \{6\} at $R$ point (Table 221 in Sec. I of the SM I). Following the conventional research approach, one would think that P-WNLs would appear at HSPs in SG 221. However, based on the previous discussion, SG 221 may not host P-WNLs at HSPs. At $\Gamma$ point, when atoms exclusively occupy any of the 1a, 1b, 3c, or 3d WYPOs, the P-WNLs will not appear. At $R$ point, when atoms exclusively occupy any of the 1a or 1b WYPOs, the P-WNLs will not appear (Table 451 in Sec. II of the SM I). When two different types of atoms occupy the 1a and 1b WYPOs, respectively, the structure still belongs to SG 221, and at this time, P-WNLs is absent. For a concrete demonstration, a couple of materials, CsCl and Na$_3$BrO, whose phonon spectra are shown in Fig. \ref{fig3} (c) and Fig. \ref{fig3} (d), are employed. In material CsCl, each of the 1a and 1b WYPOs is occupied once, so P-WNLs will not appear. While in material Na$_3$BrO, 1a, 1b, and 3c WYPOs are each occupied once, with 3c WYPO contributing once to P-WNLs at $R$ point, resulting in one occurrence of P-WNLs at the $R$ point, as indicated by the red circle in Fig. \ref{fig3} (d).

It is worth pointing out that the origin of EMP in concrete atoms can be quickly identified by our result: Once a set atoms occupy one WYPO that contributes to a special (some) EMP(s), namely,  $n_{\bm{k},j}^{w}>0$, where $\bm{k},j$ corresponds to this EMP and $w$ denotes this WYPO, one can then expect nonvanishing components of these atoms in the EMP. There are two QCTPs at the $\Gamma$ point in material CsCl, and these two QCTPs belong to the same co-irrep, which contributes from both the Cs atom and the Cl atom (Table 451 in Sec. II of the SM I). Consequently, the contribution of these two QCTPs should come from both Cs atom and Cl atom simultaneously. In contrast, at the $R$ point of the phonon spectrum of material CsCl, there are also two QCTPs. However, these two QCTPs belong to two different co-irreps contributed by Cs atom and Cl atom, respectively (Table 451 in Sec. II of the SM I). Therefore, one of these two QCTPs should come solely from the contribution of the Cl atom, and the other solely from the contribution of the Cs atom. These are all consistent with the phonon spectrum obtained by performing first-principles calculations in Fig. \ref{fig3} (c). At the $R$ point of the phonon spectrum of material Na$_3$BrO in Fig. \ref{fig3} (d), only the Na atoms occupying the WYPOs contribute to P-WNL, thus the contribution to P-WNLs comes entirely from Na atoms.

Finally, we point out that there is a triply degenerate point at the $\Gamma$ point, which is formed by the acoustic phonons regarded as Goldstone modes. This acoustic triply degenerate point is protected by the Nambu-Goldstone theorem \cite{Goldstone-1, Goldstone-2, Goldstone-3}. The detailed discussion of the representations of the triply degenerate Goldstone modes at the $\Gamma$ point is presented in the Extended Data.

\section{phonon angular momentum}
\label{sec5}

The formula for calculating the phonon AM is given by Extended Data Eq. 1. Here, we consider the AM of phonon modes at HSPs, because these phonon modes usually possess sizable intrinsic AM and well-defined phonon pseudoangular momentum \cite{Lifa-L-1,chiral-obser,phonon-computer-3}. For the AM of phonon modes at high-symmetry lines, high-symmetry planes and general positions, the approach is consistent and is left for future work. Using the phonon wavefunction obtained from symmetry analysis, we calculated the AM for all phonon modes at all HSPs in 230 SGs, and we found that the phonon modes at HSP carrying nonzero AM can only exist in 20 noncentrosymmetric SGs, as listed in Table \ref{tab3}. We determined the WYPOs which can contribute the nonzero phonon AM for the (co-)irreps at these HSPs, which are provided in Table \ref{tab3}.

Among these HSPs, the $W$ points in SG 196, SG 216 and SG 219 have six equivalent points. The directions of the phonon AM at these six points are $\pm x$, $\pm y$, and $\pm z$, respectively. The remaining HSPs each have two equivalent points, where the phonon AM is along the $z$-axis but in opposite directions. The magnitudes of the phonon AM at the equivalent points are equal.

We emphasize that, at these HSPs, there exist some WYPOs which can contribute to certain (co-)irreps, namely $n_{\bm{k},j}^{w}>0$, where $\bm{k},j$ corresponds to this (co-)irrep and $w$ denotes this WYPO, while they do not contribute to the phonon AM for this (co-)irrep. For example, when all atoms perform linear oscillation for a certain (co-)irrep, the phonon AM for this (co-)irrep is zero.

\section{CONCLUSIONS}
\label{sec6}

In summary, we demonstrated the relationship between WYPOs and phonon wavefunctions at each HSP. Firstly, we established the mapping from the WYPOs to the number of times irreps appear at each HSP in 230 SGs in the phonon spectrum. Then, we applied these results to the 111,872 materials cataloged in PhononDB@kyoto-u and ICSD, obtaining the number of times each type of EMP appears in all these materials. We identified the irreps and frequencies of each energy level at each HSP for all 10,034 materials in PhononDB@kyoto-u. Then we checked whether the EMP that can appear at HSPs will definitely be present. We discovered that among all the EMPs, only P-WNLs does not definitively exist at the corresponding HSPs in five SGs. Finally, we listed which atoms are responsible for contributing to the phonon AM. These results facilitate the control of topological phonons and phonon AM, as well as the physical phenomena associated with them.

Although there have been excellent works on phonon high-throughput calculations \cite{high-throu-phonon-2,high-throu-phonon-3}, our method and approach are different from them. In this paper, we have not only calculated the irreps for over 10,000 materials but also proposed a novel scheme that surpasses the conventional symmetry analysis method. We show that, due to the limitations of the basis functions of phonons, the EMPs which appeared can be determined based on symmetry without the need for first-principles calculations. This new methodology can be applied to other particles with Hilbert space of finite dimension, such as photons, and electrons knowing active orbitals around Fermi level. This provides a new perspective for the application of symmetry in physics.

\section{Acknowledgment}
\label{sec7}
This work was supported by the National Natural Science Foundation of China (NSFC) under Grants No. 12188101, No. 12322404, No. 12104215, No. 11834006, No. 51721001, and No. 11790311, the National Key R\&D Program of China (Grant No. 2018YFA0305704), the excellent program at Nanjing University, and Innovation Program for Quantum Science and Technology, 2021ZD0301902. F.T. was also supported by the Young Elite Scientists Sponsorship Program by the China Association for Science and Technology. X.W. also acknowledges the support from the Tencent Foundation through the XPLORER PRIZE.

\clearpage
\appendix

\section*{{Extended Data}}

\setcounter{equation}{0}
\setcounter{figure}{0}
\setcounter{table}{0}

\subsection{Details of calculation process}

For the calculation of phonon AM, the formula along the $z$ direction is as follows \cite{Lifa-L-1}:

\begin{equation}\label{am}
\renewcommand{\theequation}{Extended Data Eq. \arabic{equation}}
l^z_{{\bm{k}}\nu} = 2 \hbar  \sum_i {\rm {Im}} (\xi_{i,{\bm{k}}\nu}^{x*} \xi_{i,{\bm{k}}\nu}^{y}),
\end{equation}
where $\xi_{i,{\bm{k}}\nu}^{x}$ is the $x$-component of the phonon eigenvector with wave vector $\bm{k}$ and branch $\nu$ for the atom $i$, and the summation is performed over all atoms within the unit cell. The expressions for the phonons AM in the $y$ and $z$ directions are similar.

For the types of EMP that can appear at HSP in phonon system, the exhaustive list is as follows: charge-1 Weyl point (C-1 WP), charge-2 Weyl point (C-2 WP), charge-4 Weyl point (C-4 WP), triple point (TP), charge-2 triple point (C-2 TP), quadratic contact triple point (QCTP), Dirac point (DP), charge-2 Dirac point (C-2 DP), quadratic Dirac point (QDP), cubic crossing Dirac point (CCDP), sextuple point (SP), the nodal point located on the quadratic nodal line (P-QNL), the nodal point located on a Weyl nodal line, or at the joint point of several Weyl nodal lines that can be connected to a specific Weyl nodal line through symmetry operators (P-WNL), the nodal point located at the joint point of several Weyl nodal lines (P-WNLs), the nodal point located on a nodal surface (P-NS), the nodal point located at the intersection of two/three nodal surfaces (P-NSs), the nodal point located on a Dirac nodal line but is not situated along any high-symmetry line (P-DNL), the nodal point located at the joint point of several Dirac nodal lines (P-DNLs), and the nodal point located at the intersection of a nodal line and a nodal surface (P-WNL/NS). Among the EMPs mentioned, only the C-1 WP can occur in a general position because its existence does not typically require any specific symmetry protection, apart from lattice translation symmetry. However, its presence at a general position is accidental. The C-2 WP, TP, DP, C-2 DP, QDP, P-WNL and P-WNLs can appear at HSP or high-symmetry line, while their existence at high-symmetry line is  accidental. The rest of the EMPs can only appear at HSP. The topological properties of all these EMPs can be found in Ref. \cite{SciBul-yao}. To avoid confusion, we write the plural form of P-NS as P-NS[s], the plural form of P-NSs as P-NSss and the plural form of P-WNLs as P-WNLss in this paper.

For the collection of materials, PhononDB@kyoto-u and ICSD are the sources. For the materials obtained from PhononDB@kyoto-u, we verify their SGs and confirm that they are all consistent with those provided in the database. We also check whether the structures of these materials are under the conventions of the phonopy software \cite{phonopy}, and find that 44 materials do not, which are listed in Sec. IV of the SM I. We employ the phonopy software to standardize the structures of these 44 materials and then recalculate their phonon spectra. The calculation parameters are consistent with the original parameters in the database. For the materials from ICSD, those with duplicate and corrupt structures have been discarded. In the end, we collected all 10,034 materials from PhononDB@kyoto-u and 101,838 materials from ICSD.

Then, based on the $k \cdot p$ effective models of all types of EMPs that can appear at HSP in 230 SGs in the phonon system mentioned before \cite{SciBul-yao}, we obtain the correspondence between different types of EMPs and irreps we used at all HSPs in 230 SGs. According to the cases of atoms occupying WYPOs in the materials collected from the above two databases, we provide the number of each type of EMPs that appear at HSPs in the phonon spectra of these materials.

Next, we calculate the trace of the matrix presentations of the little group operator for the wavefunction and obtain the irreps and the type of EMP for each energy level at each HSP for 10,034 materials in the PhononDB@kyoto-u. The wavefunctions are calculated using the force constants as computed in the PhononDB@kyoto-u. The symmetry operators of 230 SGs we used are the same as those used in the phonopy software. Finally, we compare the number of times irreps appear at each HSP of 10,034 material in the PhononDB@kyoto-u, as obtained from symmetry analysis and practical first-principles calculations, and find them to be completely consistent.

\subsection{Two examples of the emergent particles enforced by Wyckoff position}

An example of subcase I in Extended Data Table 2: At the $H$ point of the SG 220, three kinds of co-irreps can appear, which are \{1, 2\}, \{3, 3\}, and \{4, 5\} (Table 220 in Sec. I of the SM I). Among them, only co-irrep \{4, 5\} corresponds to SP; nevertheless, SP will all be present when atoms occupy the 12a, 12b, 16c, 24d, or 48e WYPOs respectively (all types of WYPOs in the SG 220), appearing 2, 2, 3, 4 and 9 times, respectively (Table 220 in Sec. I and Table 450 in Sec. II of the SM I). That is to say, SP definitely exist at $H$ point in SG 220.

An example of subcase II in Extended Data Table 2: Only 1a and 1b WYPOs do not contribute to the C-4 WP at the $R$ point in SG 195 (Table 195 in Sec. I and Table 425 in Sec. II of the SM I). However, if only the 1a WYPO or 1b WYPO is occupied, or both 1a and 1b WYPOs are occupied, the resulting structure will no longer belong to SG 195. As an illustration, when an atom is only located at 1b WYPO, the structure will belong to SG 221. In this case, the C-4 WP also definitely exists at $R$ point in SG 195.

\subsection{The representations of the triply degenerate Goldstone modes}

The representations of the triply degenerate Goldstone modes at the $\Gamma$ point may consist of: (1) One three-dimensional (co-)irrep. (2) One one-dimensional (co-)irrep and one two-dimensional (co-)irrep. (3) Three one-dimensional (co-)irreps. The representation of the acoustic triply point is one three-dimensional co-irrep in the materials CsCl and Na$_3$BrO. We provide the (co-)irreps of the acoustic triply point of each SG in Sec. III of the SM I. We find that in SG 1 to SG 74, the representations of the acoustic triply point are composed of three one-dimensional (co-)irreps; in SG 75 to SG 194, the representations of the acoustic triply point are composed of one one-dimensional (co-)irrep and one two-dimensional (co-)irrep, where the two-dimensional (co-)irrep can only correspond to C-2 WP, P-QNL, P-WNL and P-WNLs; in SG 195 to SG 230, the representation of the acoustic triply point is composed of one three-dimensional (co-)irrep, which can only correspond to C-2 TP and QCTP. We also give the WYPOs which contribute to one two-dimensional (co-)irrep of the acoustic triply point in Sec. III of the SM I.

\subsection{Supplemental Material}
See Supplemental Material at \url{https://box.nju.edu.cn/d/132f8ac8aac545f2862e/}. It is divided into three parts in total: Supplemental Material I, Supplemental Material II, and Supplemental Material III. Supplemental Material I contains the irreps along with the corresponding $k \cdot p$ models and the types of EMPs for each HSP in 230 SGs; the occurrences of irreps at each HSP in phonon spectrum when atoms occupy each WYPO in 230 SGs; the irreps of the triply degenerate Goldstone modes at the $\Gamma$ point in 230 SGs; the WYPOs which contribute to one two-dimensional irrep of the triply degenerate Goldstone modes at the $\Gamma$ point in 230 SGs; the information of the 44 materials whose structure do not conform to the conventions of the phonopy software in PhononDB@kyoto-u; the information of the 2,309 nearly ideal candidate materials which we selected from the PhononDB@kyoto-u with no negative frequency phonon modes and well-separated topological phonon bands; and the coordinates of the WYPO in 230 SGs. Supplemental Material II lists the number of times the EMPs appear at each HSP in phonon spectrum in all 111,872 materials. Of these, 10,034 materials are sourced from PhononDB@kyoto-u, and the remaining 101,838 are from the ICSD. These results are obtained solely based on the crystal structure. Supplemental Material III contains the irrep, frequency, degeneracy, and the topological property of each energy level at every HSP for all 10,034 materials in the PhononDB@kyoto-u. These results are obtained based on the force constants obtained from this database. The Supplementary Material III comprises a total of five PDF files. The first four files contain detailed information on 2,000 materials each, while the fifth file includes detailed information on the remaining 2,034 materials. The sequence of the materials corresponds to the order presented in Table I of Supplementary Material II.

\renewcommand{\thetable}{\arabic{table}}

\begin{table*}[t!]
\renewcommand{\tablename}{Extended Data Table}
\caption{The EMPs enforced by representation. According to the (co-)irrep, all bands form the same type of EMP at the HSP of the corresponding SG. The number represents the SG, and the letter behind represents the HSP of this SG. The coordinates of the HSPs are listed in Sec. I of the SM I.}\label{tab1}
   \begin{tabular}{m{2.2cm}<{\centering}p{15cm}}
    \hline \hline
\noalign{\vskip 4pt}
\multirow{1}{*}{Emergent particle} & \multicolumn{1}{c}{SG and HSP}  \\ \noalign{\vskip 4pt} \hline  
\noalign{\vskip 4pt}
\multirow{1}{*}{C-1 WP}  & 24, 210: $W$; \enspace 80, 98, 199, 214: $P$ \\ 
\noalign{\vskip 4pt}
\multirow{1}{*}{DP}  & 29: $T$, $R$; \enspace 33, 60: $T$; \enspace 52: $U$; \enspace 54: $U$, $R$; \enspace 56: $U$, $T$; \enspace 73, 228: $W$; \enspace 110, 142, 206, 230: $P$; \enspace 130, 138: $R$ \\ 
\noalign{\vskip 4pt}
\multirow{1}{*}{C-2 DP}  & 19, 198, 212, 213: $R$; \enspace 92, 96: $A$ \\ 
\noalign{\vskip 4pt}
\multirow{1}{*}{QDP}  & 130, 135: $A$ \\ 
\noalign{\vskip 4pt}
\multirow{1}{*}{P-WNL}  & 39: $S$, $R$; \enspace 46, 74, 203, 227: $W$; \enspace 64: $S$; \enspace 88, 109, 122, 141: $P$; \enspace 158: $L$; \enspace 206: $N$ \\ 
\noalign{\vskip 4pt}
\multirow{1}{*}{P-WNLs}  & 7, 13: $B$, $D$, $A$, $E$; \enspace 9, 15: $A$, $M$; \enspace 14: $B$, $A$; \enspace 27, 49: $Z$, $U$, $T$, $R$; \enspace 28: $Y$, $T$, $S$, $R$; \enspace 29, 31, 53: $Y$, $S$; \enspace 30: $Y$, $Z$, $U$, $T$, $S$, $R$; \enspace 32, 50: $Y$, $X$, $U$, $T$, $S$, $R$; \enspace 33, 52: $Y$, $X$, $S$; \enspace 34, 48: $Y$, $X$, $Z$, $U$, $T$, $S$; \enspace 37, 40, 66: $Z$, $T$; \enspace 41, 68: $Z$, $T$, $S$, $R$; \enspace 43, 70: $Y$, $X$, $Z$; \enspace 45, 72: $R$, $S$, $W$; \enspace 46: $S$; \enspace 54: $X$, $S$; \enspace 56, 58, 60, 114, 135$-$138: $Z$; \enspace 67: $S$, $R$; \enspace 73: $R$, $S$, $T$; \enspace 74: $T$; \enspace 84, 105, 112, 131: $Z$, $A$; \enspace 85, 100, 104, 117, 125, 126: $M$, $A$, $R$, $X$; \enspace 86, 102, 106, 118, 133, 134: $M$, $Z$, $R$, $X$; \enspace 88, 109, 122, 141: $X$, $Z$; \enspace 101, 116, 132: $Z$, $A$, $R$; \enspace 103, 124: $R$; \enspace 108, 120, 140: $N$, $P$; \enspace 110, 142: $N$, $X$, $Z$; \enspace 158: $H$; \enspace 159, 161, 163, 165, 167, 184, 192: $L$; \enspace 188, 190: $L$, $H$; \enspace 201, 224: $X$, $M$; \enspace 203, 218, 223, 227, 228: $X$; \enspace 219, 226: $W$; \enspace 220, 230: $N$; \enspace 222: $M$ \\ 
\noalign{\vskip 4pt}
\multirow{1}{*}{P-DNL}  & 57, 60, 62: $S$, $R$; \enspace 61: $U$, $T$, $S$; \enspace 205: $M$, $R$ \\ 
\noalign{\vskip 4pt}
\multirow{1}{*}{P-DNLs}  & 61: $R$ \\ 
\noalign{\vskip 4pt}
\multirow{1}{*}{P-NS}  & 4, 11: $Z$, $C$, $D$, $E$; \enspace 14: $Z$, $C$; \enspace 17, 26, 51: $Z$, $U$, $T$, $R$; \enspace 18, 55, 57, 59: $Y$, $X$, $U$, $T$; \enspace 19, 61, 62: $Y$, $X$, $Z$; \enspace 20, 36, 63: $Z$, $T$, $R$; \enspace 29, 31, 53: $Z$, $U$; \enspace 33, 52: $Z$, $R$; \enspace 54, 64: $Z$, $T$; \enspace 56, 58, 60: $Y$, $X$; \enspace 76, 78, 91, 95: $Z$, $A$, $R$; \enspace 90, 94, 113, 114, 127, 129, 135, 137: $R$, $X$; \enspace 92, 96: $Z$, $X$; \enspace 128, 130, 136, 138, 198, 205, 212, 213: $X$; \enspace 169, 170, 173, 178, 179, 182: $A$, $L$, $H$; \enspace 176, 185, 186, 193, 194: $L$ \\ 
\noalign{\vskip 4pt}
\multirow{1}{*}{P-NSs}  & 18, 55, 56, 58, 59: $S$, $R$; \enspace 19: $U$, $T$, $S$; \enspace 62: $U$, $T$; \enspace 90, 94, 113, 127, 129, 138: $M$, $A$; \enspace 92, 96: $M$, $R$; \enspace 114, 128, 130, 135, 137, 198, 212, 213: $M$; \enspace 136: $M$, $A$, $R$ \\ 
\noalign{\vskip 4pt}
\multirow{1}{*}{P-WNL/NS}  & 14: $D$, $E$; \enspace 31, 53: $T$, $R$; \enspace 33, 60: $U$; \enspace 52: $T$; \enspace 58: $U$, $T$; \enspace 64, 128: $R$ \\ \noalign{\vskip 4pt} \hline \hline
  \end{tabular}
\end{table*}

\begin{table*}[!t]
\renewcommand{\tablename}{Extended Data Table}
\caption{The EMPs enforced by WYPO. They can be further divided into two subcases. In subcase I, at the HSP of the corresponding SG, not all bands form the same type of EMP, however, each WYPO contributes to this type of EMP. In subcase II, at the HSP of the corresponding SG, not all bands form the same type of EMP, and not each WYPO contributes to this type of EMP. However, when atoms only occupy the WYPOs that do not contribute to the EMP, the structure will belong to another SG with a higher symmetry. }\label{tab2}
   \begin{tabular}{m{0.8cm}<{\centering}m{2.2cm}<{\centering}p{14cm}}
    \hline \hline
\noalign{\vskip 4pt}
\multirow{1}{*}{Case} & \multirow{1}{*}{Emergent particle} & \multicolumn{1}{c}{SG and HSP}  \\ \noalign{\vskip 4pt} \hline
\noalign{\vskip 4pt}
\multirow{26}{*}[-34pt]{I} & \multirow{1}{*}{C-1 WP}  & 150, 152, 154, 168, 171, 172, 177, 180, 181: $K$, $H$; \enspace 169, 170, 173, 178, 179, 182: $K$ \\ 
\noalign{\vskip 4pt}
& \multirow{1}{*}{C-2 WP}  & 75, 77, 89, 93: $\Gamma$, $M$, $Z$, $A$; \enspace 76, 78, 91, 95: $\Gamma$, $M$; \enspace 79, 97: $\Gamma$, $Z$, $P$; \enspace 80, 90, 94, 98, 146, 155: $\Gamma$, $Z$; \enspace 92, 96, 169, 170, 173, 178, 179, 182: $\Gamma$; \enspace 143$-$145, 149$-$154, 168, 171, 172, 177, 180, 181: $\Gamma$, $A$; \enspace 196: $L$; \enspace 207, 208: $X$, $M$; \enspace 209: $X$, $L$, $W$; \enspace 210: $X$, $L$ \\ 
\noalign{\vskip 4pt}
& \multirow{1}{*}{C-4 WP}  & 198, 212, 213: $\Gamma$; \enspace 199, 214: $\Gamma$, $H$ \\ 
\noalign{\vskip 4pt}
& \multirow{1}{*}{TP}  & 204, 217, 229: $P$ \\ 
\noalign{\vskip 4pt}
& \multirow{1}{*}{C-2 TP}  & 195, 207, 208: $\Gamma$, $R$; \enspace 196, 198, 209, 210, 212, 213: $\Gamma$; \enspace 197, 211: $\Gamma$, $H$, $P$; \enspace 199, 214: $\Gamma$, $H$ \\ 
\noalign{\vskip 4pt}
& \multirow{1}{*}{QCTP}  & 200, 201, 215, 221, 224: $\Gamma$, $R$; \enspace 202, 203, 205, 216, 218$-$220, 222, 223, 225$-$228, 230: $\Gamma$; \enspace 204, 206, 217, 229: $\Gamma$, $H$ \\ 
\noalign{\vskip 4pt}
& \multirow{1}{*}{DP}  & 103: $Z$, $A$; \enspace 104, 161, 167: $Z$; \enspace 158, 159, 163: $A$; \enspace 165: $A$, $H$; \enspace 184, 185, 192, 193: $H$; \enspace 219, 226, 228: $L$; \enspace 220: $P$ \\ 
\noalign{\vskip 4pt}
& \multirow{1}{*}{QDP}  & 106, 114, 133, 137, 176, 184$-$186, 188, 190, 192$-$194: $A$; \enspace 124, 128: $Z$, $A$; \enspace 126, 130: $Z$; \enspace 222: $X$ \\ 
\noalign{\vskip 4pt}
& \multirow{1}{*}{CCDP}  & 220, 230: $H$ \\ 
\noalign{\vskip 4pt}
& \multirow{1}{*}{SP}  & 218, 222, 223: $R$; \enspace 220, 230: $H$ \\ 
\noalign{\vskip 4pt}
& \multirow{1}{*}{P-WNL}  & 147, 157, 175, 183, 189, 191: $K$, $H$; \enspace 159, 163: $\Gamma$, $K$, $H$; \enspace 160, 166: $\Gamma$, $Z$; \enspace 161, 167, 205: $\Gamma$; \enspace 162, 164: $\Gamma$, $A$, $K$, $H$; \enspace 165: $\Gamma$, $K$; \enspace 176, 184$-$186, 190, 192$-$194: $K$; \enspace 202, 203: $L$; \enspace 206: $\Gamma$, $H$; \enspace 216: $X$ \\ 
\noalign{\vskip 4pt}
& \multirow{1}{*}{P-WNLs}  & 87, 107, 121, 139: $P$; \enspace 103, 124: $Z$, $A$; \enspace 104, 126, 128, 130, 161, 167: $Z$; \enspace 106, 133, 159, 163, 188, 190: $A$; \enspace 147, 156$-$158: $\Gamma$, $A$; \enspace 148: $\Gamma$, $Z$; \enspace 165, 184, 192: $A$, $H$; \enspace 216, 219, 226$-$228: $L$; \enspace 220: $\Gamma$, $H$, $P$; \enspace 222: $X$; \enspace 225: $L$, $W$; \enspace 230: $\Gamma$, $H$ \\ 
\noalign{\vskip 4pt}
& \multirow{1}{*}{P-QNL}  & 81, 83, 99, 111, 115, 123: $\Gamma$, $M$, $Z$, $A$; \enspace 82, 85, 87, 100, 107, 108, 113, 117, 119, 120, 121, 125, 127, 129, 139, 140: $\Gamma$, $Z$; \enspace 84, 101, 103, 105, 112, 116, 124, 131, 132: $\Gamma$, $M$; \enspace 86, 102, 118, 134, 174, 175, 183, 187, 189, 191: $\Gamma$, $A$; \enspace 88, 104, 106, 109, 110, 114, 122, 126, 128, 130, 133, 135$-$138, 141, 142, 176, 184$-$186, 188, 190, 192$-$194: $\Gamma$; \enspace 215, 221: $X$, $M$; \enspace 218, 223: $M$; \enspace 219, 225, 226: $X$ \\ 
\noalign{\vskip 4pt}
& \multirow{1}{*}{P-NS}  & 176, 185, 186, 193, 194: $A$, $H$ \\ 
\noalign{\vskip 4pt}
& \multirow{1}{*}{P-NSs}  & 114, 128, 137: $A$ \\ 
\noalign{\vskip 4pt}
& \multirow{1}{*}{P-WNL/NS}  & 176, 186, 194: $H$ \\ \noalign{\vskip 4pt} \hline  
\noalign{\vskip 4pt}
\multirow{4}{*}[-10pt]{II} & \multirow{1}{*}{C-4 WP}  & 195, 207, 208: $\Gamma$, $R$; \enspace 196, 209, 210: $\Gamma$; \enspace 197, 211: $\Gamma$, $H$ \\ 
\noalign{\vskip 4pt}
& \multirow{1}{*}{CCDP}  & 218, 222, 223: $R$ \\ 
\noalign{\vskip 4pt}
& \multirow{1}{*}{P-WNL}  & 200, 201: $\Gamma$, $R$; \enspace 202, 203: $\Gamma$; \enspace 204: $\Gamma$, $H$, $P$ \\ 
\noalign{\vskip 4pt}
& \multirow{1}{*}{P-WNLs}  & 215, 218, 222$-$224: $\Gamma$, $R$; \enspace 217: $\Gamma$, $H$, $P$; \enspace 219, 226, 228: $\Gamma$ \\ \noalign{\vskip 4pt} \hline \hline

  \end{tabular}
\end{table*}

\renewcommand\arraystretch{1.1}\begin{table*}[!t]
\renewcommand{\tablename}{Extended Data Table}
\caption{The list of MIPD for layered materials in PhononDB@kyoto-u that can be utilized to obtain 2D materials through the exfoliation method.}\label{tab4}
   \begin{tabular}{p{1.6cm}p{1.6cm}p{1.6cm}p{1.6cm}p{1.6cm}p{1.6cm}p{1.6cm}p{1.6cm}p{1.6cm}p{1.6cm}}
    \hline \hline
mp-10009 & mp-10402 & mp-10476 & mp-10796 & mp-11131 & mp-11641 & mp-1186 & mp-12263 & mp-12307 & mp-1243 \\ 
mp-12509 & mp-13383 & mp-13384 & mp-13385 & mp-13682 & mp-13829 & mp-13923 & mp-13995 & mp-14333 & mp-1443 \\ 
mp-14791 & mp-1581 & mp-1582 & mp-16755 & mp-16819 & mp-17008 & mp-17708 & mp-17945 & mp-1821 & mp-18625 \\ 
mp-1863 & mp-1880 & mp-1943 & mp-19885 & mp-19921 & mp-20078 & mp-20331 & mp-20459 & mp-20815 & mp-20878 \\ 
mp-20902 & mp-2097 & mp-2125 & mp-21365 & mp-21405 & mp-2160 & mp-2173 & mp-2231 & mp-22323 & mp-224 \\ 
mp-2242 & mp-22577 & mp-22661 & mp-22691 & mp-22858 & mp-22877 & mp-22880 & mp-22881 & mp-22917 & mp-22939 \\ 
mp-22945 & mp-22951 & mp-22965 & mp-22968 & mp-22969 & mp-22982 & mp-22987 & mp-22993 & mp-22997 & mp-23008 \\ 
mp-23023 & mp-23024 & mp-23046 & mp-23052 & mp-23068 & mp-23072 & mp-23154 & mp-23162 & mp-23164 & mp-23170 \\ 
mp-23174 & mp-23205 & mp-23210 & mp-23218 & mp-23227 & mp-23238 & mp-23261 & mp-23293 & mp-23309 & mp-23334 \\ 
mp-23376 & mp-23395 & mp-23419 & mp-23472 & mp-23517 & mp-23553 & mp-23559 & mp-23690 & mp-23709 & mp-23803 \\ 
mp-23856 & mp-23879 & mp-23904 & mp-23918 & mp-24029 & mp-24030 & mp-24097 & mp-24120 & mp-24130 & mp-24170 \\ 
mp-2418 & mp-24204 & mp-24205 & mp-24389 & mp-24422 & mp-24428 & mp-24534 & mp-24638 & mp-2507 & mp-25469 \\ 
mp-2632 & mp-2706 & mp-27127 & mp-27164 & mp-27167 & mp-27171 & mp-27194 & mp-27195 & mp-27207 & mp-27231 \\ 
mp-27342 & mp-27356 & mp-27358 & mp-27373 & mp-27397 & mp-27411 & mp-27437 & mp-27439 & mp-27449 & mp-27455 \\ 
mp-27532 & mp-27546 & mp-27552 & mp-27622 & mp-27628 & mp-27640 & mp-27656 & mp-27666 & mp-27678 & mp-27684 \\ 
mp-27697 & mp-27700 & mp-27702 & mp-27703 & mp-27723 & mp-27734 & mp-27741 & mp-27747 & mp-27749 & mp-27770 \\ 
mp-27811 & mp-27849 & mp-27850 & mp-27863 & mp-27871 & mp-27880 & mp-27910 & mp-27922 & mp-27934 & mp-27948 \\ 
mp-27975 & mp-27976 & mp-27979 & mp-27980 & mp-2798 & mp-27998 & mp-28013 & mp-28051 & mp-28066 & mp-28067 \\ 
mp-28077 & mp-2809 & mp-28095 & mp-28109 & mp-28126 & mp-28136 & mp-28153 & mp-28178 & mp-28179 & mp-28220 \\ 
mp-28224 & mp-28253 & mp-28319 & mp-28361 & mp-28420 & mp-28448 & mp-28458 & mp-28479 & mp-28490 & mp-28491 \\ 
mp-28534 & mp-28541 & mp-28557 & mp-28569 & mp-28572 & mp-28580 & mp-28599 & mp-28627 & mp-28671 & mp-28712 \\ 
mp-28745 & mp-28809 & mp-28938 & mp-28944 & mp-29016 & mp-29027 & mp-29048 & mp-29057 & mp-29116 & mp-29117 \\ 
mp-29132 & mp-29133 & mp-29177 & mp-29218 & mp-29234 & mp-29236 & mp-29249 & mp-29251 & mp-29254 & mp-29269 \\ 
mp-29327 & mp-29355 & mp-29407 & mp-29412 & mp-29498 & mp-29509 & mp-29555 & mp-29567 & mp-29624 & mp-29666 \\ 
mp-29689 & mp-29690 & mp-29710 & mp-29798 & mp-29803 & mp-29839 & mp-30002 & mp-30031 & mp-30034 & mp-30058 \\ 
mp-3006 & mp-30096 & mp-30241 & mp-30242 & mp-30243 & mp-30244 & mp-30247 & mp-30287 & mp-30298 & mp-30299 \\ 
mp-30302 & mp-30943 & mp-30971 & mp-30993 & mp-31001 & mp-31037 & mp-31220 & mp-3123 & mp-31250 & mp-31256 \\ 
mp-31406 & mp-3228 & mp-3342 & mp-33746 & mp-341 & mp-34326 & mp-3439 & mp-3468 & mp-3502 & mp-3516 \\ 
mp-3521 & mp-36381 & mp-3779 & mp-3813 & mp-38605 & mp-3881 & mp-4468 & mp-4530 & mp-4691 & mp-4710 \\ 
mp-484 & mp-486 & mp-4888 & mp-4988 & mp-5045 & mp-504564 & mp-504630 & mp-504957 & mp-505366 & mp-505465 \\ 
mp-510128 & mp-510293 & mp-510727 & mp-5328 & mp-5338 & mp-540818 & mp-540924 & mp-541240 & mp-541487 & mp-541837 \\ 
mp-541911 & mp-541912 & mp-542096 & mp-542152 & mp-542180 & mp-542623 & mp-542812 & mp-545482 & mp-546279 & mp-546684 \\ 
mp-549720 & mp-549728 & mp-550070 & mp-550553 & mp-550820 & mp-551470 & mp-551826 & mp-552120 & mp-553025 & mp-553991 \\ 
mp-554384 & mp-554479 & mp-554921 & mp-554950 & mp-555212 & mp-555269 & mp-555490 & mp-555838 & mp-556004 & mp-556084 \\ 
mp-556094 & mp-556216 & mp-556231 & mp-556516 & mp-556523 & mp-557163 & mp-557250 & mp-557437 & mp-557614 & mp-557685 \\ 
mp-557835 & mp-557992 & mp-558672 & mp-559065 & mp-559208 & mp-559332 & mp-559379 & mp-559760 & mp-560196 & mp-560370 \\ 
mp-560665 & mp-560845 & mp-560847 & mp-560923 & mp-561011 & mp-561190 & mp-561489 & mp-5625 & mp-567279 & mp-567441 \\ 
mp-567471 & mp-567556 & mp-567720 & mp-567798 & mp-567809 & mp-567931 & mp-567949 & mp-568252 & mp-568301 & mp-568328 \\ 
mp-568346 & mp-568592 & mp-568949 & mp-568971 & mp-569126 & mp-569224 & mp-569687 & mp-569766 & mp-570084 & mp-570125 \\ 
mp-570136 & mp-570157 & mp-570172 & mp-570219 & mp-570321 & mp-570340 & mp-570418 & mp-570590 & mp-570610 & mp-570823 \\ 
mp-570930 & mp-570961 & mp-571169 & mp-571219 & mp-571260 & mp-571442 & mp-571550 & mp-573376 & mp-574169 & mp-580886 \\ 
mp-580941 & mp-5962 & mp-602 & mp-6023 & mp-604315 & mp-620029 & mp-6215 & mp-623087 & mp-624190 & mp-625054 \\ 
mp-625150 & mp-625513 & mp-625521 & mp-625526 & mp-625541 & mp-625548 & mp-625630 & mp-626497 & mp-626550 & mp-626557 \\ 
mp-6300 & mp-630329 & mp-630528 & mp-632760 & mp-634859 & mp-642725 & mp-642807 & mp-643359 & mp-643708 & mp-643759 \\ 
mp-643896 & mp-644222 & mp-644223 & mp-665 & mp-667 & mp-6684 & mp-672273 & mp-674328 & mp-675543 & mp-675651 \\ 
mp-676020 & mp-676250 & mp-676572 & mp-691 & mp-696175 & mp-696944 & mp-696946 & mp-696949 & mp-698205 & mp-700 \\ 
mp-7008 & mp-7090 & mp-720 & mp-7280 & mp-7293 & mp-7310 & mp-7461 & mp-7505 & mp-752460 & mp-752637 \\ 
mp-753173 & mp-753309 & mp-753425 & mp-753445 & mp-753575 & mp-753670 & mp-753712 & mp-753720 & mp-754031 & mp-754113 \\ 
mp-754152 & mp-754217 & mp-754246 & mp-754300 & mp-754329 & mp-754494 & mp-754526 & mp-754815 & mp-754969 & mp-755336 \\ 
mp-755350 & mp-755463 & mp-755565 & mp-755689 & mp-755703 & mp-755705 & mp-756437 & mp-756578 & mp-756702 & mp-757145 \\ 
mp-759405 & mp-759883 & mp-765493 & mp-768906 & mp-769028 & mp-769077 & mp-770718 & mp-772282 & mp-772295 & mp-7784 \\ 
mp-779704 & mp-780260 & mp-7868 & mp-7991 & mp-8140 & mp-8147 & mp-8285 & mp-8299 & mp-8300 & mp-8612 \\ 
mp-8616 & mp-8946 & mp-8976 & mp-9006 & mp-9010 & mp-9018 & mp-909 & mp-9254 & mp-9272 & mp-9321 \\ 
mp-938 & mp-9479 & mp-9509 & mp-9518 & mp-9548 & mp-9622 & mp-9657 & mp-972032 & mp-972117 & mp-973585 \\ 
mp-973981 & mp-975666 & mp-976349 & mp-976713 & mp-977371 & mp-978553 & mp-9815 & mp-983327 & mp-985829 & mp-985831 \\ 
mp-9889 & mp-989189 & mp-989192 & mp-989193 & mp-989196 & mp-9897 & mp-9912 & mp-9921 & mp-9922 \\
\hline \hline
  \end{tabular}
\end{table*}

\clearpage

\bibliography{references}
\end{document}